\documentclass[prx,twocolumn,english,superscriptaddress,floatfix,longbibliography]{revtex4-2}

\usepackage{subcaption}
\usepackage{graphicx}
\usepackage{amsmath}
\usepackage{amssymb}
\usepackage{amsfonts}
\usepackage{dsfont}
\usepackage{dcolumn}
\usepackage{bbm}
\usepackage{ragged2e}
\usepackage{mathtools}
\usepackage{braket}
\usepackage{physics}
\usepackage{siunitx}
\usepackage[unicode]{hyperref}
\usepackage[nameinlink]{cleveref}
\Crefname{section}{Sec.}{Secs.}
\Crefname{equation}{Eq.}{Eqs.}
\Crefname{figure}{Fig.}{Figs.}
\Crefname{tabular}{Tab.}{Tabs.}

\usepackage{bbold} 
\usepackage{nicefrac}
\usepackage{tikz}
\usetikzlibrary{arrows.meta}
\usepackage{lipsum}  
\usepackage{xcolor}

\usepackage[makeroom]{cancel}

\usepackage{xfrac}

\usepackage[linesnumbered,ruled,vlined]{algorithm2e}
\usepackage{orcidlink}
\usepackage[normalem]{ulem}
\definecolor{tdo_blue}{HTML}{0000FF}
\definecolor{tdo_darkgreen}{HTML}{839A00}
\definecolor{tdo_red}{HTML}{FF0000}

\newcommand{\bes}{\begin{subequations}}
\newcommand{\ees}{\end{subequations}}
\newcommand{\be}{\begin{equation}}
\newcommand{\ee}{\end{equation}}

\makeatletter
  \newcommand\flausr{\@fleqntrue}
\makeatother

\begin{document}
    \title{Emergent-Coupling-Based Ansatz Evaluated on a Superconducting Quantum Processor}


    \author{Alina Joch\,\orcidlink{0009-0001-9391-6653}}
    \affiliation{Condensed Matter Theory, TU Dortmund University, Otto-Hahn-Stra\ss{}e 4, 44227 Dortmund, Germany}
    \affiliation{Institute of Software Technology, German Aerospace Center (DLR), 51147 Cologne, Germany}

    \author{Kevin Lively\,\orcidlink{0000-0003-2098-1494}}
    \affiliation{Institute of Software Technology, German Aerospace Center (DLR), 51147 Cologne, Germany}
    
    \author{Benedikt Fauseweh\,\orcidlink{0000-0002-4861-7101}}
    \email[Corresponding author: ]{benedikt.fauseweh@tu-dortmund.de}
    \affiliation{Condensed Matter Theory, TU Dortmund University, Otto-Hahn-Stra\ss{}e 4, 44227 Dortmund, Germany}
    \affiliation{Institute of Software Technology, German Aerospace Center (DLR), 51147 Cologne, Germany}

    \date{\textrm{\today}}

\begin{abstract}
The performance of the variational quantum eigensolver depends critically on the choice of ansatz. In this work, we experimentally evaluate the emergent-coupling-based ansatz (ECBA), a physically motivated variational ansatz for disordered systems. The ECBA is based on a renormalization (semi-)group approach to determine the dominant effective couplings \textcolor{black}{and construct an asymptotically correct ground state. A variational layer based on the location and strength of the effective couplings then corrects the limitations of the renormalization group approach while preserving} shallow circuits that capture the essential long-range entanglement structure and balance local correlations. We implement the ECBA on superconducting quantum processors and benchmark it on disordered Heisenberg chain models. Using classically pre-optimized parameters and error mitigation techniques, we study systems of up to 30 qubits and observe an experimental relative energy accuracy of 96.4{1}\% for the largest system. Furthermore, we find that the ECBA can be efficiently embedded on hardware with two-dimensional square-lattice connectivity. We compare to commonly used hardware efficient ansätze and observe that the ECBA achieves significantly higher accuracy at a similar gate count.
\end{abstract}

\maketitle

\section{Introduction}

Simulating quantum many-body systems that are beyond the reach of classical methods \cite{feynm82, lloyd96} is a primary motivation for the development of quantum computers. This capability is of particular relevance for the study of complex materials, chemical reactions, and strongly correlated quantum systems \cite{bulut09, brown10, georg14,fause24}. However, current quantum hardware operates in the noisy intermediate-scale quantum (NISQ) regime \cite{knill98, presk18}, where limited qubit numbers, finite coherence times, and gate errors severely constrain the depth and complexity of executable quantum circuits. These limitations pose significant challenges for achieving accurate and scalable quantum simulations, emphasizing the crucial importance of shallow and resource-efficient quantum circuits.

Among the most studied approaches for exploiting NISQ devices is the Variational Quantum Eigensolver (VQE) \cite{cerez21, peruz14}, a hybrid classical–quantum algorithm that approximates ground-state properties of quantum systems using parameterized quantum circuits. In VQE, a classical optimizer iteratively updates circuit parameters to minimize an energy expectation value measured on a quantum device. The practical performance of VQE is strongly influenced by hardware noise, optimization difficulties, and the choice of ansatz \cite{Kandala2017}. In particular, issues such as barren plateaus can severely hinder convergence, especially for deep or highly expressive circuits \cite{McClean2018,Cerezo2021,Arrasmith2021effectofbarren}. A barren plateau is a regime in which the variational energy landscape becomes exponentially flat with system size, causing gradients to vanish and making variational optimization extremely difficult due to the large measurement  necessary to resolve meaningful gradient estimates. Consequently, the design of efficient, hardware-compatible ansätze remains a central challenge in VQE-based quantum simulation \cite{lyu20, tkach21, Zhang2022,PhysRevResearch.6.043254}.

Various strategies have been proposed to improve ansatz performance \cite{TILLY20221}. Hardware-efficient ansätze (HEA) \cite{Kandala2017} are tailored to the native gate sets and connectivity of quantum devices, thereby reducing circuit depth \textcolor{black}{ and facilitating implementation on near-term hardware. However,} they often exhibit limited problem-specific expressivity, can suffer from barren plateaus, \textcolor{black}{and may require substantial optimization effort to accurately approximate complex many-body ground states.
To overcome these limitations, a broad range of physically motivated ansätze has been developed.} Problem-inspired approaches such as the Variational Hamiltonian Ansatz \cite{wecke15} aim to mimic adiabatic state preparation, potentially improving convergence, although at the cost of reduced flexibility or increased circuit depth. 
\textcolor{black}{Similarly, chemistry-inspired ansätze such as UCCSD encode physically motivated excitation operators \cite{barko18, romer19, abhin22, jiaqi25}, while adaptive approaches dynamically construct  circuits during the optimization. Beyond these approaches, symmetry-preserving circuits exploit conserved quantities to reduce the accessible Hilbert space and improve trainability \cite{gard20, burto24}.} More recently, approaches such as the Quantum-optimal-control-inspired ansatz \cite{choqu21} have explored symmetry breaking and control-theoretic concepts to enhance expressivity and optimization performance.

Despite these advances, identifying ansätze that simultaneously balance expressivity, trainability, \textcolor{black}{hardware efficiency,} and noise resilience remains an open problem, particularly for disordered systems\textcolor{black}{, where rugged energy landscapes and frustration complicate state preparation} \cite{PhysRevB.107.024204,Cao2025exploitingmanybody}. Systems with long-range interactions and disorder represent a driver in the study of strongly correlated systems, due to the complexity of their correlations, such as rare region effects \cite{Vojta_2006}, and the challenges they pose for classical simulation methods \cite{pared05, sanch10, vojta19, defen23, buchh23, kovac22, Wilming2023, lin23, kovac24}.

\textcolor{black}{In this work we investigate, how an advanced many-body method, strong disorder renormalization-group (SDRG), can be used to derive an effective circuit ansatz and how this ansatz can be implemented efficiently without overhead onto a state-of-the-art superconducting quantum processor. This emergent-coupling-based ansatz (ECBA) uses information from SDRG to identify the relevant initial state and couplings in the parameterized quantum circuit \cite{Joch_2025}. The central idea of ECBA is that, while in principle every qubit may interact with every other qubit, only a subset of couplings contributes significantly to the relevant low-energy physics. }

\textcolor{black}{Hierarchical renormalization-based methods such as the multi-scale entanglement renormalization ansatz (MERA) have demonstrated the potential of exploiting scale-dependent entanglement structures for efficient quantum state representations \cite{vidal7, berez25}. In contrast to MERA, which provides a general tensor-network representation optimized to describe many-body quantum states, the ECBA uses renormalization-group information to construct a problem-specific variational quantum circuit tailored to the dominant couplings of disordered systems.} 
This approach yields shallower circuits while retaining the essential entanglement structure of the target states and improved performance over naive variational ansätze. Such shallow circuits enable error mitigation techniques to be applied efficiently.
We implement and assess the ECBA on a superconducting quantum processor. Benchmarking against a generic HEA, we show that the ECBA accurately represents ground states of systems with up to 30 qubits while the HEA does not achieve comparable accuracy. We further realize an embedding of the ECBA on square lattice hardware and show that the long-range entanglement structure identified by a renormalization group approach can be resolved experimentally.


This paper is organized as follows. In Sec.~\ref{chap:methods}, we introduce the methods used in our simulations, including a detailed description of the error mitigation techniques employed, and an overview of the employed quantum devices. The paper proceeds with a case study of the rainbow chain model as an introductory {and preliminary} example in Sec.~\ref{chap:rainbow}, followed by the more challenging random quantum critical chain model in Sec.~\ref{chap:rg}. For both systems, we discuss the choice of ansatz, the embedding on the hardware, and a detailed comparison of the results obtained with different ansätze. We conclude our analysis in Sec.~\ref{sec:conc}.

\section{Methods}
\label{chap:methods}

The VQE implementation is described in detail in Appendix \ref{vqe}. In this work, we use a pre-optimized set of parameters obtained from the Python tensor-network package quimb \cite{quimb}, meaning that the classical optimization steps have already been performed to determine the best parameters for the chosen quantum circuit. Numerical details can be found in Appendix \ref{numericaldetails} and code is available in Ref.~\cite{joch26}. We stress that the present work is concerned primarily with the expressivity of the ansatz, and not with its optimizability. Because the circuits studied here are predominantly shallow, optimization is not expected to be a limiting factor. This is confirmed in Sec.~\ref{chap:rg}, where classical simulations show that the variational parameters can be optimized reliably even for relatively large numbers of qubits.

\subsection{Error Mitigation} 

Quantum Error Mitigation (EM) is a rapidly evolving area of quantum computing which, in contrast to error correction, does not attempt to fix errors in the evolution of an encoded state through active measurement and feedback. Instead, EM focuses on mitigating the effects of noise on specific expectation values through averaging over several modified versions of the logical  circuit in combination with pre- and post-processing \cite{EM,filippov2024scalabilityquantumerrormitigation}. In this study, we employ several such techniques. 

First, we use Pauli Twirling \cite{twirling1, twirling2} to mitigate coherent errors of two-qubit entangling operations by ``dressing'' each two-qubit gate in four random single-qubit Paulis, such that it remains logically unchanged. This reduces the impact of coherent noise while reducing the worst-case error rate.
To address readout errors, we utilize Twirled Readout Error Extinction (TREX) \cite{trex}. This technique diagonalizes the readout error map by applying random bit-flip operations just before measurement, followed by inversion in post-processing. This method significantly improves the fidelity of the measurement process, which is crucial for accurate results. We average over 16 TREXed and twirled circuit instantiations, each consisting of 10,000 shots, to ensure statistical robustness. 

We furthermore implemented dynamical decoupling \cite{dd}, which reduces unwanted crosstalk, dephasing, and decoherence by applying a sequence of single-qubit gates to idling qubits, followed by their inverses. We use a minimal sequence of two ${\text{Pauli-}X}$ gates during idle sequences.
Finally, we apply digital Zero Noise Extrapolation (ZNE) \cite{zne}. This method manually boosts the noise present in the circuit through the application of logically redundant two-qubit gates, which usually account for the majority of noise on superconducting devices. By undoing and redoing the entangling operations everywhere they appear, the base noise level, $\lambda = 1$, associated with the circuit can be amplified in increments as $\lambda\in 2\mathbb{N}+1$. An extrapolation to the zero-noise level $\lambda = 0$ can then be made, using an exponential noise model $O(\lambda)=O_0 \, \mathrm{exp}(a\lambda)$. These error mitigation strategies were implemented manually on top of Qiskit's ``Sampler'' primitive \cite{quantumcomputingqiskit}.

{ In practice, we calculate every observable reported below as follows: for a given circuit instantiation where the two-qubit gates have been dressed, and random bit-flip operations applied, we calculate the observable over the 10,000 samples. We then average these expectation values over the 16 instantiations, carrying through uncertainty from the shot noise from each instantiation by adding in quadrature. We perform this for $\lambda\in \{1, 3, 5\}$, for a total of 48 application circuits. The error bars reported below come from a Levenberg-Marquardt fitting procedure where the above calculated shot noise is propagated through as uncertainty. The dominant factor in the uncertainty comes from the fitting procedure and not the shot noise.}

\subsection{Quantum Processors}
We use two quantum computing devices provided by IQM, the Garnet device with 20 qubits and the Emerald device with 54 qubits. Both devices are based on superconducting transmon qubits, which are widely used in quantum computing due to their relatively long coherence times and ease of manipulation. The qubits in both devices are arranged in a square lattice configuration and are connected via tunable couplers, allowing for control of qubit interactions. The systems are calibrated to support Phased X Rotation (PRX) gates, which implement arbitrary X and Y rotations as native single-qubit gates (i.e., rotations of the form $\exp[-i\frac{\theta}{2}(\cos\phi X+\sin\phi Y)]$) and CZ as the native two-qubit gate \cite{meetiqm, abdur24}. Some important hardware properties are listed in Tab.~\ref{tab:iqmdevices}.
\begin{table}[htb]
\centering
    \begin{tabular}{l|c c c c}
        device  & T$_1$ & T$_2$ & PRX fidelity & CZ fidelity \\ \hline 
         Garnet  & 36.53 $\mathrm{\mu s}$ & 8.61 $\mathrm{\mu s}$ & 99.91 \% & 99.37 \% \\
        Emerald & 50.05 $\mathrm{\mu s}$ & 15.75 $\mathrm{\mu s}$ & 99.94 \% &  99.45 \%
    \end{tabular}
    \caption{\label{tab:iqmdevices} \justifying Median values of the T1 relaxation time, T2 coherence time (Ramsey), PRX gate fidelity, and CZ gate fidelity for the IQM Garnet and IQM Emerald devices \cite{meetiqm,abdur24}.}
\end{table}

Qubit selection and circuit transpilation are performed using the Qiskit transpile function, based on a square lattice embedding. This ensures that the circuits are properly optimized for the hardware architecture of the respective device, without requiring any SWAP gates.
For the measurement process, we perform readout in the $x$, $y$, and $z$ bases, as required for calculating the energy expectation value of the specific Hamiltonians under investigation in this work.

\section{Rainbow chain}
\label{chap:rainbow}

As an introductory example, we consider the rainbow chain \cite{Ramirez_2015, Alba_2019}, described by the Hamiltonian
\begin{equation}\label{eq:sl}
    H_{\mathrm{RC}} = \alpha \sum_{i=1}^{{n-1}} \left( \vec{S}_i \cdot \vec{S}_{i+1} \right) + J  \sum\limits_{i=1}^{n/2} \left(\vec{S}_i \cdot \vec{S}_{n-(i-1)} \right),
\end{equation}
where $n$ denotes an even total number of qubits.
This one-dimensional chain model can be mapped to a qubit ladder, see Fig.~\ref{fig:model_rainbow}, in which the long-range couplings are mapped to the rungs of the ladder. 
We set $J/\alpha = 100$, i.e. with dominant long-range interactions, and analyze the effect of explicitly incorporating long-range connections into the ansatz, as opposed to restricting it to short-range terms.
In the rainbow-chain model, the long-range interactions are already present at the Hamiltonian level, so the dominant coupling structure is manifest from the outset.

\subsection{Circuit and Embedding for the Rainbow Chain}
\begin{figure*}[htb]
        \centering
        \begin{subfigure}[t]{0.32\linewidth}
            \includegraphics[width=\linewidth]{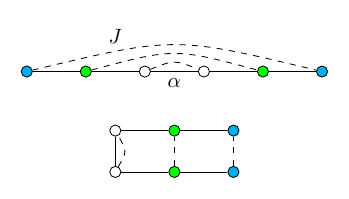}
            \vspace{-1em}
            \caption{}
            \label{fig:model_rainbow}
        \end{subfigure}
        \hspace{0.3em}
        \vspace{1em} 
        \begin{subfigure}[t]{0.22\linewidth}
            \includegraphics[width=\linewidth]{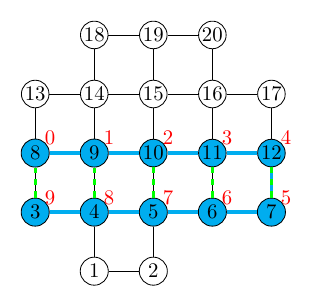}
            \vspace{-1em}
            \caption{}
            \label{fig:embedding_rainbow}
        \end{subfigure}
        \hspace{0.3em}
        \vspace{1em}
        \begin{subfigure}[t]{0.37\linewidth}
            \includegraphics[width=\linewidth]{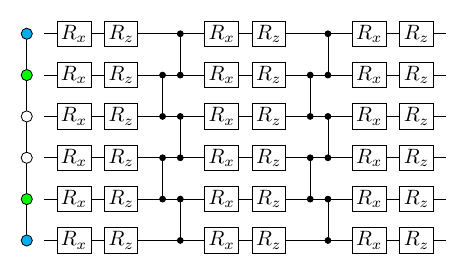}
            \vspace{-1em}
            \caption{}
            \label{fig:ansatze_rainbow_HEA}
        \end{subfigure}
        \begin{subfigure}[t]{0.49\linewidth}
            \vspace{-1.5em}
            \includegraphics[width=\linewidth]{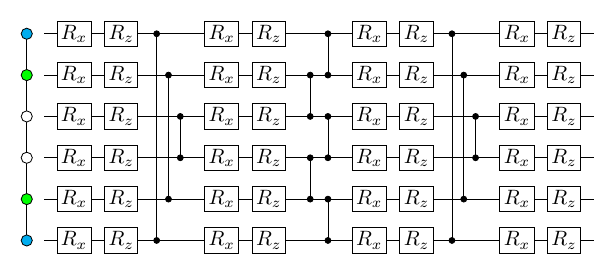}
            \vspace{-1em}
            \caption{}
            \label{fig:ansatz_rainbow_CBA}
        \end{subfigure}
        \vspace{1em}
        \begin{subfigure}[t]{0.45\linewidth}
            \vspace{-1.5em}
            \includegraphics[width=0.8\linewidth]{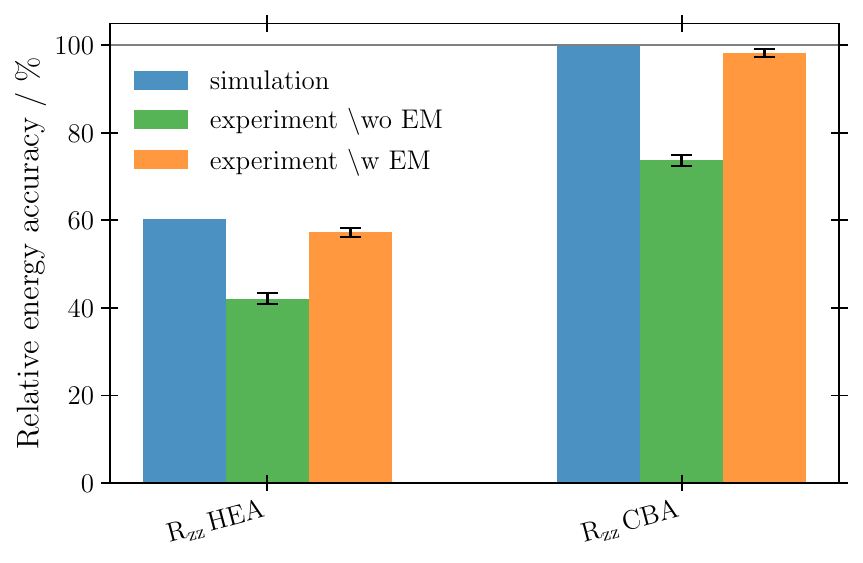}
            \vspace{-0.5em}
            \caption{}
            \label{fig:results_rainbow}
        \end{subfigure}
    \vspace{-1.5em}
    \caption{\justifying 
        (a) Schematic illustration of the rainbow model for a system size of six qubits. The upper part shows the qubit chain, the lower part a mapping to a qubit ladder structure.
        (b) Example of an embedding of the rainbow chain for the CBA onto the IQM Garnet device. Blue connection lines indicate nearest-neighbor gates, green dashed lines denote long-range gates, and red numbers label the mapping of the $i$-th qubit of the chain to the corresponding physical qubit.
        (c) Hardware-efficient ansatz (HEA) consisting exclusively of single-qubit and nearest-neighbor gates, shown as an example for six qubits.
        (d) Coupling-based ansatz (CBA) composed of alternating layers of single-qubit gates combined with long-range two-qubit gates reflecting the coupling structure of the model and layers of single-qubit gates combined with nearest-neighbor gates.
        (e) Relative accuracy of the simulated (noise-free) data and the experimental data obtained for 10 qubits on the IQM Garnet device, each compared to the exact value, for both the HEA and the CBA with and without error mitigation. Error mitigation is applied only to the experimental data.
    }
\end{figure*}

We compare different ansätze, each pre-optimized by classical simulation. 
We evaluate a linear hardware-efficient ansatz (HEA) consisting of two hardware-efficient layers in a brick wall structure, and a coupling-based ansatz (CBA) consisting of one layer including the long-range couplings of the system, one nearest-neighbor layer and a second long-range layer. One $R_x$ and one $R_z$ gate are added to each qubit at the end of the circuit in both cases. The ansätze are depicted in Fig.~\ref{fig:ansatze_rainbow_HEA} and Fig.~\ref{fig:ansatz_rainbow_CBA}, respectively.
\textcolor{black}{
Note that, by construction, the HEA with its two-layer architecture can generate correlations only up to a maximum range of six spins. This inherent limitation already suggests that the ansatz will encounter difficulties in representing the ground state of larger problem instances, where longer-range correlations become essential. We show that there exists the possibility to create an ansatz with the same circuit depth as the HEA that is nevertheless capable of realizing the required correlations and accurately capturing the ground state.
}

Strictly speaking, our HEA does not meet the definition of hardware efficiency established in Ref.~\cite{Kandala2017} because we omit ladder-rung coupling gates. This, however, is exactly the point we want to illustrate here. Our HEA labeled method attempts to force a one-dimensional layout on a system that clearly shows long-range couplings, which turns out to be far less efficient than including the natural entanglement geometry.

Table~\ref{tab:numbersansatze_rainbow} compares the number of gates, the circuit depth, and the number of variational parameters for the two ansätze for 10 qubits.

\begin{table}[htb]
\centering
    \begin{tabular}{l|c c c}
        ansatz & \# CZ gates \,\,  & circuit depth \,\,   & \# parameters \\ \hline 
        HEA  & 36 & 18 & 78 \\
        CBA  & 38 & 18 & 99 
    \end{tabular}
    \caption{\label{tab:numbersansatze_rainbow} \justifying Comparison of number of two-qubit gates, depth of the circuit, and number of variational parameters for the HEA and the CBA for 10 qubits after transpilation for the IQM Garnet device.}
\end{table}

For accurate results, all qubit pairs acted on by two-qubit gates should be adjacent on the square lattice of the quantum processor, because nonlocal operations require additional SWAP gates and thereby lead to substantial errors.
For the HEA, this is given by construction as two-qubit gates are only applied to neighboring qubits. The CBA, however, includes two-qubit gates between non-neighboring qubits. In the case of the rainbow chain, the embedding is straightforward by mapping the chain to a qubit ladder structure. This is visualized in Fig.~\ref{fig:embedding_rainbow} for 10 qubits on the IQM Garnet device.

\subsection{Results for the Rainbow Chain}

We perform simulations on a 10-qubit rainbow chain model. First, we consider ideal, noise-free VQE outcomes, which we refer to as simulation results, throughout the remainder of this work. This provides a baseline to determine the maximum accuracy the chosen ansatz can achieve on an error-free quantum device.
We then evaluate the energy on the real device, where noise is present, and refer to this data as experimental results. In the case of the experimental results, we distinguish between calculations performed with and without error mitigation. For each circuit, 10,000 shots are used. Consequently, without error mitigation a total of 30,000 shots are performed, as we need one circuit for each measurement basis. In the case where error mitigation is applied, each individual circuit contributing to the mitigated value is also executed with 10,000 shots. In total, for the rainbow chain and for the random Heisenberg chain, we used 1,920,000 shots. 

Figure~\ref{fig:results_rainbow} evaluates the performance of the HEA and the CBA. Specifically, we plot the relative accuracy of both the experimental and simulation data with respect to the numerically computed exact solution, defined as
\begin{align}
 \epsilon_{\mathrm{sim, exp}} = 1 - \frac{| x_{\mathrm{exact}} - x_{\mathrm{sim, exp}}|}{| x_{\mathrm{exact}}|} \, ,\   
\end{align} 
where $x_{\mathrm{exact}}$ denotes the exact value, and $x_{\mathrm{exp}}$ and $x_{\mathrm{sim}}$ refer to the values obtained on the real device (experiment) and the ideal noise-free VQE (simulation), respectively.

While the CBA simulation yields highly accurate results with a relative error of $10^{-9}$, the ideal HEA simulation reaches an relative energy accuracy of only $\sim 60\%$. In Ref.~\cite{Joch_2025} we show that the required number of HEA layers needed to reach accuracies comparable to those of a fixed depth CBA scales with the system size. 
However, on the real quantum device, this increase in circuit depth amplifies the impact of gate errors and decoherence, leading to lower overall fidelity. The CBA achieves comparable accuracy with fewer layers, thereby mitigating the adverse effects of hardware noise.
Even without error mitigation, we find that the CBA outperforms the HEA. 
Nonetheless, error mitigation is needed for experimental data to reproduce the exact value. With the CBA and error mitigation, we obtain $98.20\%\pm 0.95\%$ relative energy accuracy.

\textcolor{black}{
Our results clearly demonstrate that the CBA significantly outperforms the conventional method. It should be noted, however, that the simulations presented here were carried out for a comparatively large ratio of $J/\alpha$. As shown in Ref.~\cite{Joch_2025}, the accuracy of the CBA decreases as this ratio is reduced, with the CBA outperforming the conventional approach starting from $J/\alpha = 1$. To further validate our findings under these conditions, we complemented the experiments performed on the real IQM quantum device with additional simulated experiments using the IQM FakeBackend at $J/\alpha = 1$. 
Using the simulated experimental data, we obtain a relative accuracy of $80.13\%\pm1.49\%$ with respect to the exact results for the HEA, whereas the CBA achieves a relative accuracy of $90.04\%\pm0.74\%$. These results demonstrate that, even in a regime where the simulated advantage is less pronounced, the CBA still yields more accurate results when executed on a noisy quantum device.
}

{The rainbow chain serves as a preliminary result of how long-range couplings can be handled in a VQE ansatz. We now use this knowledge to build an ansatz for a model in which long-range couplings are not explicit in the Hamiltonian but emerge from the underlying interactions. We therefore refer to this ansatz as the emergent CBA (ECBA). The underlying idea remains the same.}

\section{Random quantum critical point model}
\label{chap:rg}

In random quantum critical qubit chains, quenched randomness drives the system towards the random-singlet phase, a critical state in which qubits pair into singlets over broadly distributed length scales rather than only between neighboring sites \cite{PhysRevLett.43.1434,PhysRevB.22.1305,refae04}. In this regime, the pattern of singlet bonds is not imposed microscopically, but emerges due to bond fluctuations, leading to long-range entanglement across the entire chain \cite{refae04}. We consider the random Heisenberg chain
\begin{equation}
    H_{\mathrm{RH}} = \sum_{i=1}^{{n-1}} J_i \vec{S}_i \cdot \vec{S}_{i+1} \, ,
\end{equation}
where the number of qubits $n$ is chosen to be even and the couplings $J_i > 0$ are drawn from any non-singular distribution. 
In our case, we use the distribution
\begin{equation}\label{eq:distribution}
    P(J) = \frac{1}{\delta} J^{-1+\delta^{-1}} 
\end{equation}
with a disorder strength $\delta \geq 1$, as proposed in Ref.\@ \cite{laflo05}. Larger values of $\delta$ correspond to broader coupling distributions and hence stronger disorder. We study several disorder realizations for different values of $\delta$ and system sizes $n$.

The formation of nonlocal singlets can be understood within the strong disorder renormalization-group (RG) \cite{Igloi2018} framework. 
Iteratively, the strongest bond $J_i$ satisfying $J_i \gg J_{i-1}, J_{i+1}$ is 
identified, and a singlet is built between the two qubits coupled by this bond. 
The neighboring qubits of the resulting singlet are then coupled by the smaller
renormalized coupling strength \cite{refae04}
\begin{equation}
    J'_{i-1, i+2} = \frac{J_{i-1} J_{i+1}}{2 J_i} \, 
\end{equation}
arising from second-order perturbation theory.
Repeating these two steps until all qubits are paired into singlets yields an accurate approximation of the ground state of the system, with singlets connecting arbitrarily distant sites. This description becomes asymptotically exact in the limit of strong disorder and large distances.
In this way, an initial state can be constructed entirely from the classical RG flow, without prior optimization on the quantum device.   
\textcolor{black}{We use this initial state as an approximate starting point, which we refine using a variational layer, in which the two qubit-gates are balancing long-range correlations derived from the RG flow and local correlations. The algorithm to obtain the ansatz is provided in Appendix~\ref{pseudocode}.}

\subsection{Circuit and Embedding for the Random Quantum Critical Point Model}
\begin{figure*}[htb]
      \centering
      \hspace{-6em}
      \begin{subfigure}[b]{0.48\textwidth}
          \includegraphics[width=0.73\linewidth]{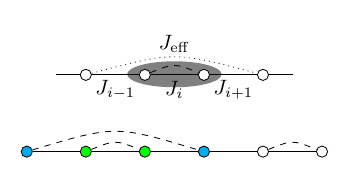}
          \vspace{-0.7em}
          \caption[]{}
          \vspace{2.5em}
          \label{fig:model_rg}
     \end{subfigure}
     \hspace{-0em}
     \begin{subfigure}[t]{0.26\textwidth}
          \includegraphics[width=0.85\linewidth]{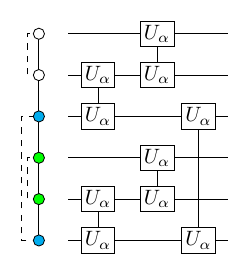}
          \vspace{-0.5em}
          \caption[]{}
          \label{fig:ansatz_rg_ECBA}
     \end{subfigure}
     \hspace{1em}
     \begin{subfigure}[t]{0.2\textwidth} 
          \includegraphics[width=0.85\linewidth]{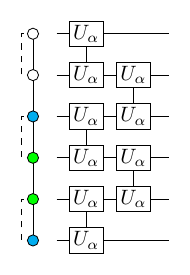}
          \vspace{-0.5em}
          \caption[]{}
          \label{fig:ansatz_rg_HEA}
     \end{subfigure}

    \begin{tikzpicture}[remember picture,overlay]
        \node[align=center] at (-2.1,4.65) {\small circuit design (b) \\ based on RG flow (a)};
        \draw[->,thick,red]
            (-3,4)
            .. controls (-2.5,4.3) and (-1.5,4.3) ..
            (-1,4);
    \end{tikzpicture}

     \vspace{-4em}
     \begin{subfigure}[t]{0.34\textwidth}
          \includegraphics[width=\linewidth]{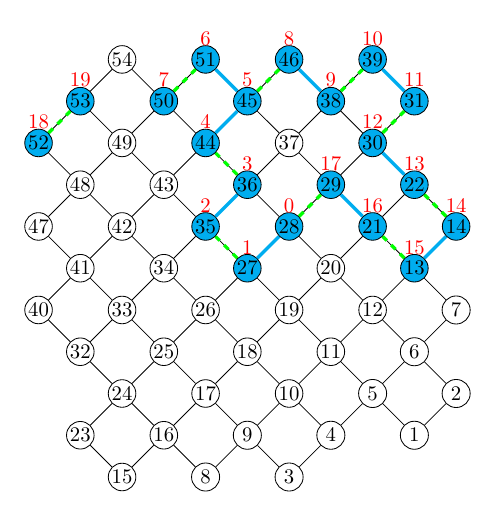}
          \vspace{-1.5em}
          \caption[]{}
          \label{fig:embedding_rg}
     \end{subfigure}
     \begin{subfigure}[t]{0.65\textwidth}
          \includegraphics[width=\linewidth]{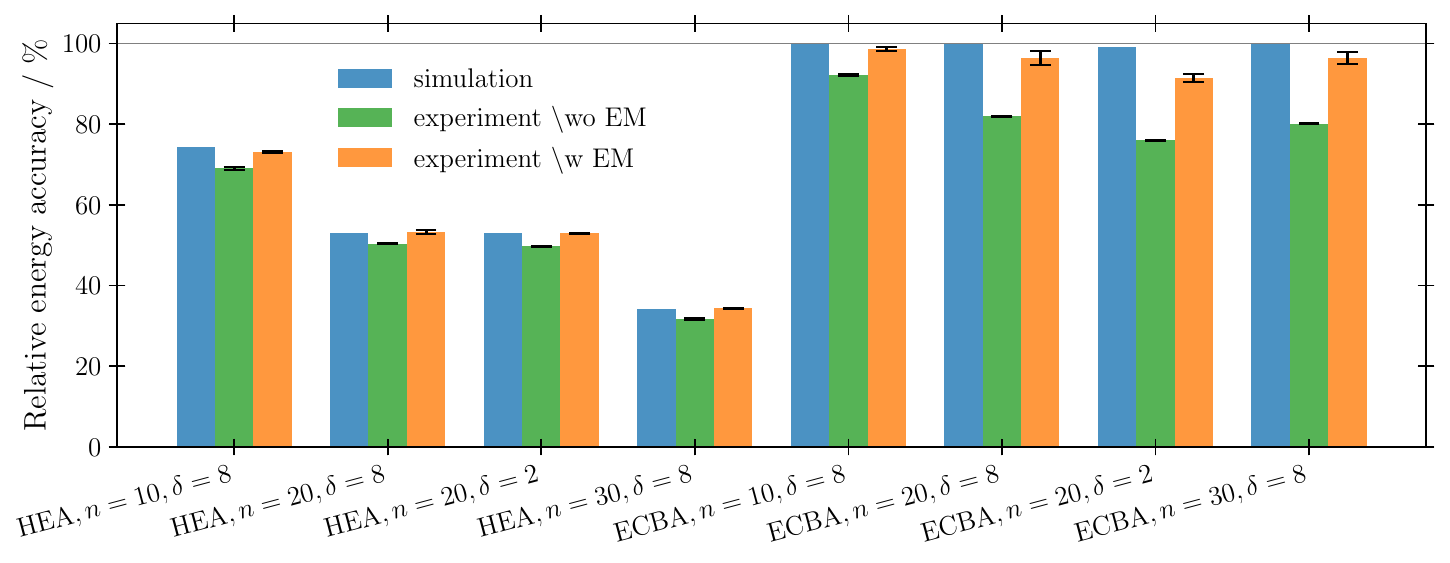}
          \vspace{-1.5em}
          \caption[]{}
          \label{fig:results_rg}
     \end{subfigure}
     \caption{\justifying 
        (a) Schematic illustration of the random quantum critical point model. The upper part depicts the RG decimation procedure: if $J_i$ is the strongest coupling in the chain, sites $i$ and $i+1$ form a singlet (dashed line). Quantum fluctuations generate an effective interaction between sites $i-1$ and $i+2$ (dotted line). The lower part shows one possible configuration illustratively for six qubits, where singlets may connect arbitrarily distant sites.
        (b) ECBA for the random chain, shown for six qubits and a single realization of the RG flow. The initial state is constructed from singlets corresponding to the RG singlet pairs. The variational circuit first applies $U_\alpha$ gates to qubit pairs associated with the strongest couplings not forming RG singlets, followed by $U_\alpha$ gates acting on the RG singlet pairs.
        (c) HEA for the random chain, shown for six qubits: The initial state is constructed from singlets between neighboring qubits (dashed lines), with solid lines indicating the $J$ couplings, followed by a layer of $U_\alpha := R_{xx} R_{yy} R_{zz}$ gates entangling the qubits in a brick wall structure.
        (d) Embedding of the 20 qubit configuration with disorder parameter $\delta=2$, used for the results shown in Figs.~\ref{fig:results_rg}, \ref{fig:heatmap} and \ref{fig:heatmap2}, on the IQM Emerald device. Green dashed lines indicate RG singlet pairs, blue lines denote the strongest remaining couplings not associated with RG singlets, and red numbers label the mapping from logical qubit indices to physical qubits. The long-range singlet pairs are given by (5, 8) and (0, 17). The two isolated singlets (18,19) emerge because the influence of the long-range couplings outweighs that of the couplings between the neighboring qubits (17,18).
        (e) Relative energy accuracy of experimental results {without and} with EM and of ideal simulated results for different system sizes and disorder parameters, each referenced to the exact ground-state energy.
     }
\end{figure*}
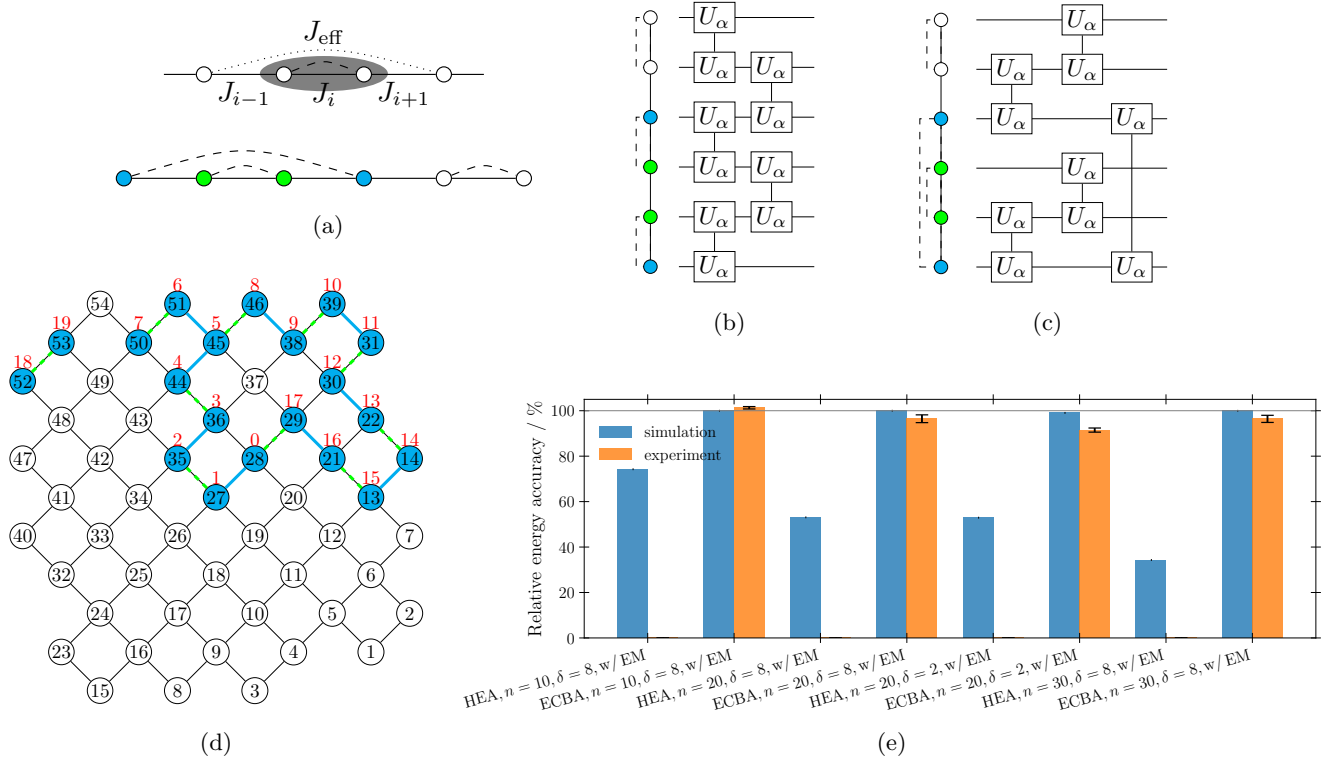

For the HEA, we initialize the system by forming singlets between nearest neighbors and thus in the $S_\text{tot}=0$ symmetry sector. We then apply $U_\alpha := R_{xx} R_{yy} R_{zz}$ gates in a brick wall structure, potentially allowing the ansatz to stay within the symmetry sector \cite{Crognaletti_2025}. The total number of $U_\alpha$ gates is $n-1$.
In the ECBA, we incorporate the RG singlets into the structure of the circuit to capture the dominant couplings. First, the singlet initial state is formed according to the long-range interactions resulting from the RG flow, then we apply the variational circuit, which consists of two parts.
In the latter part, $n/2$ $U_\alpha$ gates are applied to the qubits forming singlets according to the RG flow. Reference~\cite{Joch_2025} showed that the performance of the ansatz improves further when RG singlets are incorporated not only in the initial state but also in the variational circuit itself.
The former part consists of applying $U_\alpha$ gates to neighboring qubit pairs with the strongest couplings, excluding those already coupled within the RG flow. This is done until we reach a number of $n/2-1$ gates. This procedure allows us to balance the effects of both short- and long-range couplings in the variational circuit.
{In Appendix~\ref{appendix:ECBAconstruction}, we analyze the improvements introduced at each step of the ECBA. This allows to distinguish the improvements resulting from the RG initial state from those arising from the variational circuit selecting the strongest coupling.}
Using {the ECBA}, we obtain the same total number of gates and parameters as in the HEA, see the 30-qubit case in Tab.~\ref{tab:numbersansatze_rg}, allowing for a direct comparison. Note that consequently, in the ECBA, gates are not applied between all nearest neighbors, as some local neighbor gates are replaced by long-range gates targeting RG singlets.
The ansätze are depicted in Fig.~\ref{fig:ansatz_rg_ECBA} and Fig.~\ref{fig:ansatz_rg_HEA}, respectively.

\begin{table}[htb]
\centering
    \begin{tabular}{l|c c c}
        ansatz & \# CZ gates \,\,  & circuit depth \,\,   & \# parameters \\ \hline 
        HEA  & 189 & 27 & 87 \\
        ECBA  & 189 & 39 & 87 
    \end{tabular}
    \caption{\label{tab:numbersansatze_rg} \justifying Comparison of number of two-qubit gates, depth of the circuit, and number of variational parameters for the HEA and the ECBA for 30 qubits after transpiling them on the IQM Emerald device.}
\end{table}

When laying out the circuit onto the hardware, we want all qubit pairs acted on by two-qubit gates to be adjacent. Since the singlet pairs arise randomly from the RG flow, the mapping is not necessarily trivial as it was previously in the case of the rainbow chain. 
The problem corresponds to a unit embedding \cite{alegr24, jaxin19} of a planar graph with a maximum degree of three and no triangles. While it is not trivial to show analytically that an embedding is always possible, numerical tests suggest that it is feasible in practice. 
\textcolor{black}{
To test this, we implemented a recursive backtracking algorithm that incrementally constructs a self-avoiding walk on the square lattice while enforcing the chain connectivity and the nearest-neighbor constraints imposed by the singlet pairs. Whenever a partial embedding violates one of these constraints, the algorithm backtracks and explores an alternative branch of the search tree.
}
For systems with 20 qubits, we sampled 1,000,000 random configurations of RG singlet pairs, and in all cases a valid embedding was found. Similarly, for systems with 40 qubits, a valid embedding was obtained for all 1,000 randomly generated configurations. For these configurations, all nearest-neighbor interactions were included alongside the long-range couplings. In the case of the ECBA not all nearest-neighbor couplings are present, meaning that the graph is no longer connected, which further increases the likelihood of a feasible embedding. We therefore conclude that, except for potentially rare cases, a valid embedding exists for each RG configuration provided that the employed quantum device is sufficiently large.

We perform experiments for a system size of 10, 20, and 30 qubits. For 10 qubits, we use the IQM Garnet device, and for 20, and 30 qubits, the larger IQM Emerald device. For example, an embedding of 20 qubits corresponding to a sampled disorder realization with $\delta=2$ can be seen in Fig.~\ref{fig:embedding_rg}.
\textcolor{black}{The values of the couplings for the different configurations used are given in Ref.~\cite{joch26}.}

\begin{figure}[htb] 
    \includegraphics[width=\linewidth]{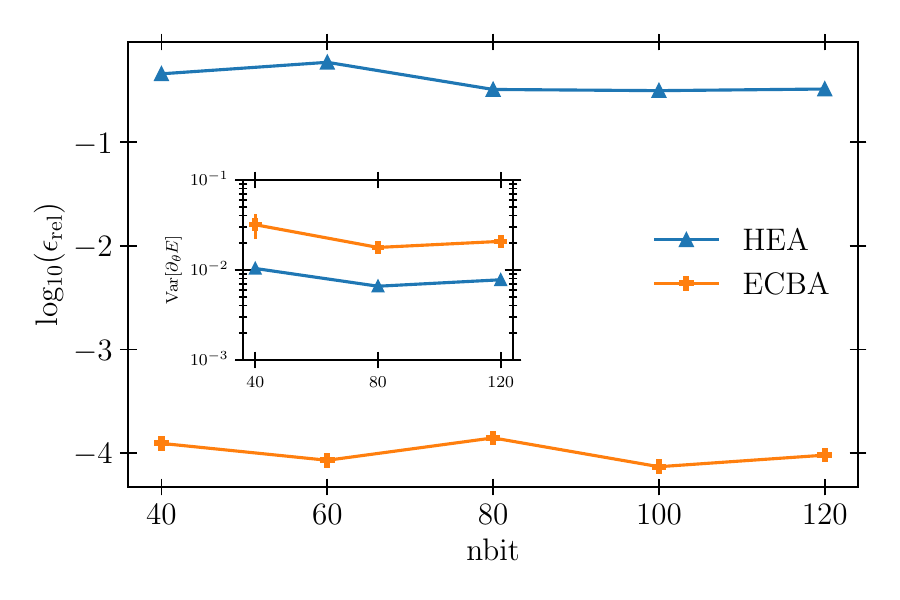}
    \vspace{-0.5em}
    \caption[]{\justifying  Logarithmic value of the relative error between the respective VQE simulation and the exact result as function of the system size. The inset shows the average variance of the gradient sampled over {100 initial parameter sets each}. The error bars indicate the variance across the different gradient components.}
    \label{fig:scaling}
\end{figure}

\begin{figure}[htb]
    \includegraphics[width=\linewidth]{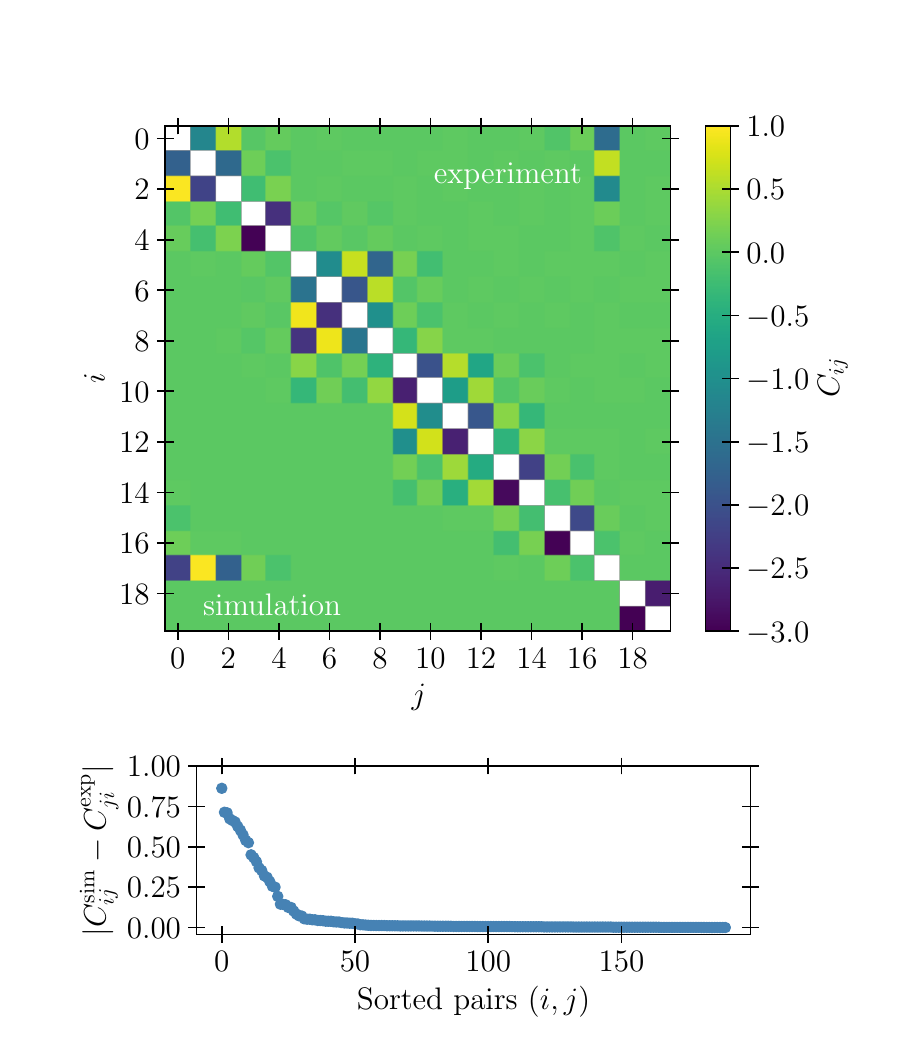}
    \vspace{-0.5em}
    \caption[]{\justifying  Correlation matrix $C_{ij}$ for 20 qubits and $\delta=2$ for the ECBA. Comparison between ideal simulated results of the VQE circuit (lower left half) and experimental results (upper right half). {The bottom panel sorts the entries by the absolute error between simulation and experiment.}}
    \label{fig:heatmap}
\end{figure}

\begin{figure}[htb] 
    \includegraphics[width=\linewidth]{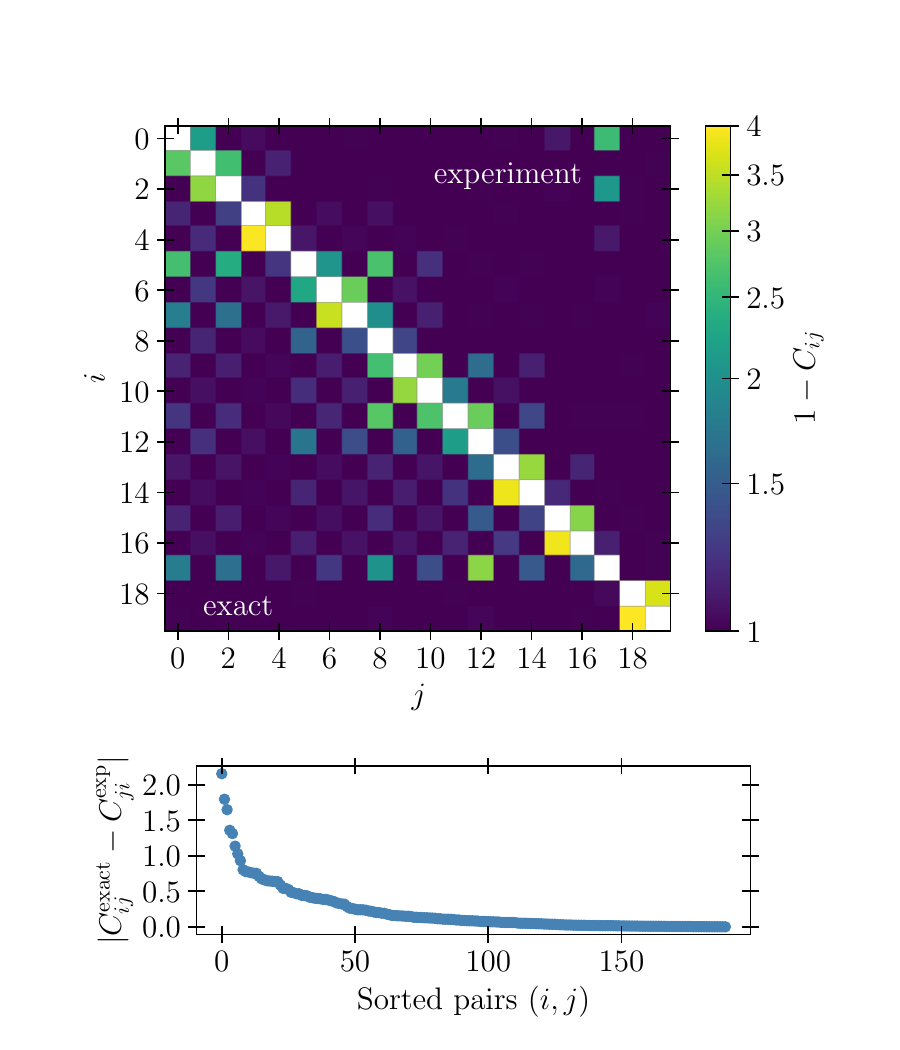}
    \vspace{-0.5em}
    \caption[]{\justifying  Correlation matrix $1 - C_{ij}$ for 20 qubits and $\delta=2$ for the ECBA shown on a logarithmic scale. Comparison between exact (lower left half) and experimental results (upper right half). {The bottom panel sorts the entries by the absolute error between experiment and exact calculation.}}
    \label{fig:heatmap2}
\end{figure}

\subsection{Results for the Random Quantum Critical Point Model}
We compute the optimal energy with ECBA for different system sizes and different disorder strengths $\delta$ and compare the results with the exact value. As before, we use pre-optimized circuits and 10,000 shots per circuit for the experimental results.
We first consider a system size of 10 qubits and a disorder strength of $\delta=8$. The experimental result obtained with the ECBA on the quantum processor  corresponds to {$98.69\%\pm0.52\%$} of the exact value. 
To investigate the scalability of the approach, we increase the system size to 20 and 30 qubits while keeping $\delta=8$. In these cases, the experimental result reaches $96.47\%\pm1.71\%$ and $96.41\%\pm1.57\%$ of the exact result, respectively. 
As shown in Ref.~\cite{Joch_2025}, variational circuits utilizing an RG structure achieve excellent accuracy, especially at strong disorder. The relative error between the ECBA simulation and the exact solution at $\delta=8$ is only $\sim10^{-4}$.

For smaller $\delta$, however, the relative error of the ECBA simulation compared to the exact result increases. Therefore, in a final step we perform the calculation for 20 qubits and $\delta=2$ to investigate how the accuracy depends on disorder strength. We obtain $91.48\%\pm0.91\%$ agreement of the experimental result with the exact result. As expected, the accuracy is reduced for smaller $\delta$, but the result remains close to the exact solution.
Figure~\ref{fig:results_rg} presents these results, demonstrating also that the HEA is significantly less accurate than the ECBA. 

For the HEA, it is known that in critical systems the number of layers, and thus the circuit depth, required to achieve high accuracy scales with system size \cite{bravo20}. In particular, once a certain system size is exceeded, a single-layer ansatz no longer provides accurate results. This behavior is evident in Fig.~\ref{fig:results_rg}, where the optimized HEA performs poorly for one layer.
In contrast, the ECBA performs well already at a single-layer depth. To examine the dependence on system size, we plot the logarithm of the relative error as a function of system size in Fig.~\ref{fig:scaling} for $\delta=8$ for simulations of up to 120 qubits. The data show that the ECBA outperforms the HEA by approximately four orders of magnitude across the entire range of system sizes considered, without degradation as the system size increases. Because critical ground states are generally difficult to represent with shallow local circuits, whose finite depth induces an effective finite correlation length, the key observation here is that the ECBA reaches a substantially higher accuracy than the HEA, consistent with the inclusion of the relevant nonlocal correlations

To further examine the impact of barren plateaus, we show in the inset of Fig.~\ref{fig:scaling} the variance of the energy gradient as a function of system size sampled over a large number of parameter initializations. A larger variance is beneficial for optimization, as nonzero gradient components provide directions along which the parameters can be updated. Previous works have shown that the gradient variance typically approaches zero as both the circuit depth and the system size increase, leading to the onset of barren plateaus \cite{McClean2018, holme22}. If one of these quantities is kept fixed, the variance is expected to converge to a finite value. This behavior is visible in our setup, where only a single layer is employed. We also find that the ECBA exhibits a slightly larger gradient variance than the HEA, indicating a somewhat more favorable optimization landscape already at minimal depth. At the same time, achieving comparable accuracy with the HEA would require a circuit depth that grows with system size, which in turn increases the effect of noise and barren plateaus. Note that noise does not lead to barren plateaus for shallow circuits \cite{Wang2021}.

To visualize the presence of long-range entanglement, we investigate the correlation matrix  $C_{ij} = \langle X_iX_j \rangle + \langle Y_iY_j \rangle +\langle Z_iZ_j \rangle$, for 20 qubits and $\delta=2$. 	Note that a single entry serves as an entanglement witness for the corresponding qubit pair when $C_{ij} < -1$. 
In Fig.~\ref{fig:heatmap}, we show this matrix for the experimental data obtained with the ECBA and for the VQE simulation {together with the absolute difference of those two sorted by size of the difference}. The heatmap clearly shows that the experimental data closely match the simulated results. The off-diagonal  correlations coincide with the long-range RG singlet pairs. {The sorted differences show that only a few pairs exhibit larger differences, whereas the majority show little to no substantial difference.}
In Fig.~\ref{fig:heatmap2}, we compare $C_{ij}$ between the experimental data of the ECBA to the exact calculation {and again the sorted absolute difference}. We observe that the exact heatmap shows richer structure, with more values deviating from zero. The experimental results accurately capture the structures around the RG singlets, reproducing both short- and long-range correlations very well. As expected, correlations at intermediate ranges are missing, which can be attributed to the limited range of our approach, restricted to a single layer.

Overall, the ECBA maintains robust agreement with exact calculations for systems up to 30 qubits, even in the more challenging regime of weak disorder. In contrast the performance of the HEA degrades with system size. This suggests that the HEA is unable to capture the long-range correlations in this random quantum critical model for a single layer. This underlines the role of the ansatz, as a well-suited variational circuit is essential for achieving high accuracy. 

\section{Conclusions}
\label{sec:conc}

In this work, we investigate the emergent-coupling-based ansatz (ECBA) for variational quantum simulations on NISQ devices. We motivate its construction in the rainbow chain model, where the relevant long-range couplings are explicitly encoded in the Hamiltonian. This example allowed us to directly investigate how building the variational ansatz around the dominant couplings leads to shallow circuits that accurately reproduce the ground-state properties on superconducting quantum hardware, achieving {an relative energy accuracy} of up to 98\% after error mitigation. Building on this controlled setting, we then applied the ECBA to disordered quantum critical qubit chains, where the relevant correlations emerge only implicitly through the renormalization flow.

Across both models, the ECBA consistently outperforms commonly used hardware-efficient ansätze (HEA) by reaching higher accuracy at a similar gate count. Our numerical simulations show that the renormalization-based structure of the ECBA enables an efficient representation of long-range entanglement in strongly disordered systems, and remains effective for system sizes up to the simulated 120 qubits, suggesting potential scalability beyond the experimental regime explored here. 

We further argue that the ECBA can be efficiently embedded on superconducting quantum processors with two-dimensional square lattice connectivity, making it compatible with state-of-the-art hardware architectures. The resulting reduction in circuit depth mitigates the impact of decoherence and gate errors and allows error mitigation techniques to be applied effectively, thereby extending the accessible system sizes on present-day devices.

Beyond the specific qubit-chain models studied here, a generalization of the ECBA provides a framework for constructing variational circuits based on emergent couplings in the strong disorder renormalization group sense. 
Natural extensions include {other disordered models} \cite{Igloi2018}, higher-dimensional lattice systems \cite{kovac10, chepu23}, applications in quantum chemistry \cite{Kandala2017}, and non-equilibrium dynamics \cite{Fauseweh2021,Fauseweh2023quantumcomputing} where identifying dominant effective couplings may similarly enable shallow yet expressive ansätze. 
{Examples for other potentially interesting models for the ECBA method are the disordered transverse-field Ising model \cite{refae04}, the Ashkin-Teller model \cite{Igloi2018} and the random Hubbard model \cite{igloi06}. The disordered transverse-field Ising model is a particularly interesting candidate for an extension to higher dimensions, as it retains an infinite disorder fixed point in these dimensions \cite{kovac22, kovac24}.  
When considering quantum chemistry applications, it would be interesting to investigate the performance of our ansatz for molecular systems \cite{omall16, Kandala2017, Masso15}. Quantum chemistry problems are often addressed using VQE with hardware-efficient ansätze or chemically inspired ansätze such as UCCSD \cite{Lolur2023, naseri25, bistafa25}. However, finding improved ansätze remains an open challenge.
For non-equilibrium dynamics, the effective Floquet hamiltonian could be interesting as it can yield a dimerized structure \cite{Katz2025}. For example, there exist one-dimensional periodically driven Floquet systems which are controlled by infinite-randomness fixed points of a SDRG flow \cite{Berda2018}, disorder induced critical phases in periodically driven dimerized Kitaev chains \cite{Roy2024}, and symmetry-enriched random quantum spin chains \cite{duque21}.
}
More broadly, our results suggest that leveraging emergent coupling structures provides a systematic route to shallow circuit initialization for disordered eigenstate problems.\\

\begin{acknowledgments} 
This work has been supported by the Quantum Fellowship Program of the German Aerospace Center (DLR). We acknowledge IQM for providing access to their quantum computing resources and for their technical support.
This work was funded by the Deutsche Forschungsgemeinschaft (DFG, German Research Foundation)—Project No. FA
1884/5-1. Q-Neko project has received funding from the European Union’s Horizon Europe research and innovation programme under Grant Agreement No. 101241875. This work was also performed for Council for Science, Technology and Innovation (CSTI), Cross-ministerial Strategic Innovation Promotion Program (SIP), “Promoting the application of advanced quantum technology platforms to social issues”(Funding agency: QST).
We further thank E. Lobe for helpful discussions.
\end{acknowledgments}

\bibliography{sample.bib}

\onecolumngrid

\begin{appendix}
\section{VQE}
\label{vqe}
The Variational Quantum Eigensolver (VQE) is a quantum-classical hybrid algorithm designed to approximate the ground state of a Hamiltonian $H$ for a system of $n$ qubits. In the VQE, a parameterized quantum circuit is constructed, described by the unitary operator $U(\theta)$. The goal of the VQE is to optimize the parameters $\theta$ in such a way that the resulting quantum state $U(\theta) |0\rangle^{\otimes n}$ provides the best approximation of the ground state of the Hamiltonian.

The quality of this ansatz is quantified by the energy expectation value$ E(\boldsymbol{\theta}) = \langle \psi(\boldsymbol{\theta}) | H | \psi(\boldsymbol{\theta}) \rangle.$ In VQE, this cost function is evaluated on the quantum device, while a classical optimizer updates the parameters $\boldsymbol{\theta}$ to minimize $E(\boldsymbol{\theta})$ in an iterative manner. The resulting optimal parameters, $\boldsymbol{\theta}_{\mathrm{opt}}$, define the variational approximation to the ground state,
$|\psi(\boldsymbol{\theta}_{\mathrm{opt}})\rangle = U(\boldsymbol{\theta}_{\mathrm{opt}}) |0\rangle^{\otimes n},
$
with corresponding energy $E(\boldsymbol{\theta}_{\mathrm{opt}})$, which provides an upper bound to the exact ground-state energy.

\section{Numerical Details}
\label{numericaldetails}
For the variational optimization of the tensor network state, we employ the \texttt{TNOptimizer} class from the quimb library, implemented in Python. We use the L-BFGS-B algorithm as the underlying optimizer and rely on JAX as the automatic differentiation backend. The variational parameters are initialized as small random numbers drawn uniformly in the range $[0,0.05]$. The optimization is performed until convergence is reached, with a tolerance of $\mathrm{tol} = 10^{-6}$ on the change of the loss function, while other algorithmic parameters are kept at the default settings of \texttt{scipy.optimize.minimize}.

Exact reference values are obtained via Density matrix renormalization group (DMRG) using quimb. The Hamiltonian is represented as a matrix product operator, and the DMRG bond dimensions are increased progressively up to 600. The DMRG calculations use a cutoff of $10^{-12}$, a convergence tolerance of $10^{-5}$, and a maximum of 150 sweeps.

\section{RG-Based Selection of Variational Couplings}
\label{pseudocode}

The placement of the variational two-qubit gates $U_\alpha$ follows a two-stage procedure: (i) a RG selection of singlet pairs, and (ii) an augmentation by the strongest remaining couplings without renormalization.

\begin{algorithm}[H]
\SetAlgoLined
\KwIn{Number of qubits $N$, random couplings $J$}
\KwOut{List of singlet pairs $P_\mathrm{RG}$}
Initialize empty list $P_\mathrm{RG}$\;
$J_{\mathrm{temp}} \gets J$\;
$I \gets [0,1,\dots,N-1]$\;

\While{$I \neq \emptyset$}{
    $k \gets \arg\max_i J_{\mathrm{temp},i}$\;
    
    \If{$|I| == 2$}{ 
        Add pair $(I_0,I_1)$ to $P_\mathrm{RG}$\;
        Remove entries $0$ and $1$ from $I$ and entry $0$ from $J_{\mathrm{temp}}$\;
    }
    \ElseIf{$k == 0$}{ 
        Add pair $(I_0, I_1)$ to $P_\mathrm{RG}$\;
        Remove entries $k$ and $k+1$ from $I$ and $J_{\mathrm{temp}}$\;
    }
    \ElseIf{$k == |J_{\mathrm{temp}}|-1$}{ 
        Add pair $(I_k, I_{k+1})$ to $P_\mathrm{RG}$\;
        Remove entries $k$ and $k+1$ from $I$ and $J_{\mathrm{temp}}$\;
    }
    \Else{ 
        Renormalize neighboring couplings: $J_{\mathrm{temp}, k-1} \gets \frac{J_{\mathrm{temp},k-1} \cdot J_{\mathrm{temp},k+1}}{2 J_{\mathrm{temp},k}}$\;
        Add pair $(I_k, I_{k+1})$ to $P_\mathrm{RG}$\;
        Remove entries $k$ and $k+1$ from $I$ and $J_{\mathrm{temp}}$\;
    }
}
\Return{$P_\mathrm{RG}$}\;
\caption{RG pairing of spins}
\end{algorithm}

\begin{algorithm}[H]
\SetAlgoLined
\KwIn{Number of qubits $N$, random couplings $J$}
\KwOut{List of added pairs $P_\mathrm{add}$}
Compute $P_\mathrm{RG}$ with Algorithm 1\;
$J_{\mathrm{temp}} \gets J$\;
Initialize empty list $P_\mathrm{add}$\;
\While{$|P_{\mathrm{add}}| < N/2 - 1$}{
    $k \gets \arg\max_i J_{\mathrm{temp},i}$\;
    \eIf{pair $(k,k+1) \in P_\mathrm{RG}$}{
        $J_{\mathrm{temp},k} \gets 0$\;
    }{
        Add $(k,k+1)$ to $P_\mathrm{add}$\;
        $J_{\mathrm{temp},k} \gets 0$\;
    }
}
\Return{$P_\mathrm{add}$}\;
\caption{Selection of strongest remaining couplings}
\end{algorithm}

\begin{algorithm}[H]
\SetAlgoLined
\KwIn{Parameter vector $\theta$, pairs $(P_\mathrm{RG}, P_\mathrm{add})$}
Initialize parameter iterator over $\theta$\;
\For{each pair $(i,j)$ in $P_\mathrm{add}$}{
    Apply $U_\alpha(i,j;\theta)$\;
}
\For{each pair $(i,j)$ in $P_\mathrm{RG}$}{
    Apply $U_\alpha(i,j;\theta)$\;
}
\caption{Construction of the variational circuit}
\end{algorithm}

\section{Construction of the ECBA}
\label{appendix:ECBAconstruction}

\begin{figure}[htb] 
    \includegraphics[width=0.5\linewidth]{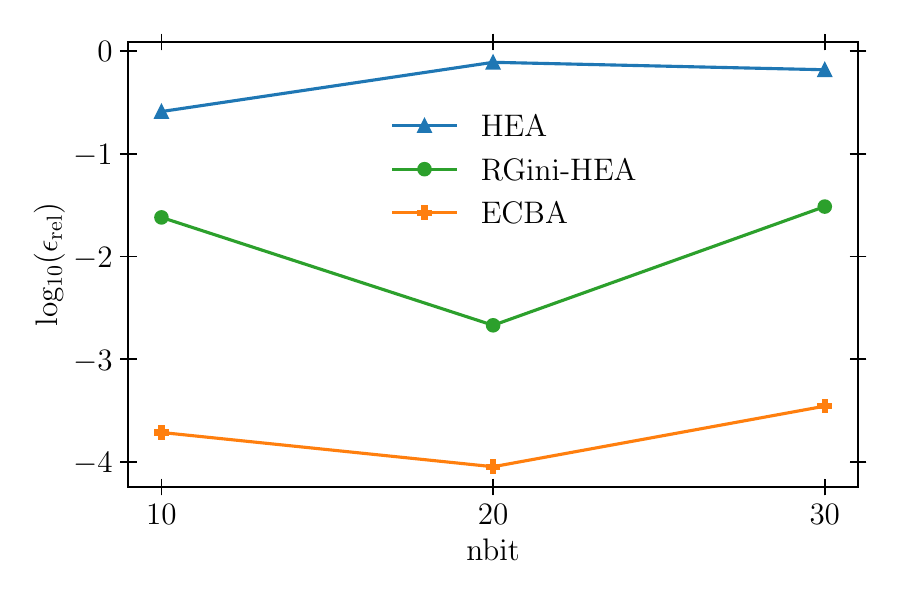}
    \vspace{-0.5em}
    \caption[]{\justifying  { Logarithmic value of the relative error between the respective VQE simulation and the exact result as function of the system size for the HEA, the ECBA, and a combination where the RG initial state is used together with the variational part of the HEA. The three systems considered here are the three systems used for the IQM experiments in Fig.~\ref{fig:results_rg} for $\delta=8$. } }
    \label{fig:app_scaling}
\end{figure}

{
Here, we take a closer look at how the ECBA is constructed and how it operates. To this end, we plot in Fig.~\ref{fig:app_scaling} the logarithmic value of the relative error for the three systems considered in the IQM experiments in Fig.~\ref{fig:results_rg} for $\delta=8$. In addition to the HEA and the ECBA, we consider an intermediate ansatz that uses the RG state as its initial state, followed by the HEA variational circuit.}

{
One can see that adapting the initial state to the RG structure already leads to an improvement of approximately one to two orders of magnitude. Incorporating this structure into the variational circuit together with the strongest couplings, i.e., using the ECBA, results in a further improvement of about one to three orders of magnitude. This leads to an overall improvement of approximately four orders of magnitude for all systems.
This behavior is consistent with the results presented in Ref.~\cite{Joch_2025}, where similar behavior was observed for a system size of 120.
}

\end{appendix}

\end{document}